\documentclass[aip,onecolumn,preprint]{revtex4-1}
\usepackage{graphicx}
\DeclareGraphicsExtensions{.pdf,.jpg,.png,.eps}
\usepackage{amsfonts}
\usepackage{amsmath}
\usepackage{amssymb}
\usepackage{color}
\usepackage{color}
\usepackage{epstopdf}

\begin{document}
\title{Double-phase transition and giant positive magnetoresistance in the quasi-skutterudite Gd$_3$Ir$_4$Sn$_{13}$}
%
%
%
\author{Harikrishnan S. Nair}
\email{h.nair.kris@gmail.com}
\affiliation{Highly Correlated Matter Research Group, Physics Department, P. O. Box 524, University of Johannesburg, Auckland Park 2006, South Africa}
\author{Sarit K. Ghosh$^\ddagger$}
\affiliation{Highly Correlated Matter Research Group, Physics Department, P. O. Box 524, University of Johannesburg, Auckland Park 2006, South Africa}
\affiliation{Department of Applied Physics, Birla Institute of Technology, Mesra 835215, Ranchi, Jharkhand, India}
\author{Ramesh Kumar K.}
\affiliation{Highly Correlated Matter Research Group, Physics Department, P. O. Box 524, University of Johannesburg, Auckland Park 2006, South Africa}
\author{Andr\'{e} M. Strydom}
\affiliation{Highly Correlated Matter Research Group, Physics Department, P. O. Box 524, University of Johannesburg, Auckland Park 2006, South Africa}
\affiliation{Institute of Physics, Chinese Academy of Sciences, PO Box 623, Beijing 100190, China}
\affiliation{Max Planck Institute for Chemical Physics of Solids (MPICPfS), N\"{o}thnitzerstra{\ss}e 40, 01187 Dresden, Germany}
\date{\today}
\begin{abstract}
The magnetic, thermodynamic and electrical/thermal transport properties of the caged-structure quasi-skutterudite Gd$_3$Ir$_4$Sn$_{13}$ are re-investigated. The magnetization $M(T)$, specific heat $C_p(T)$ and the resistivity $\rho(T)$ reveal a double-phase transition -- at $T_{N1}\sim$ 10~K and at $T_{N2}\sim$ 8.8~K -- which was not observed in the previous report on this compound. The antiferromagnetic transition is also visible in the thermal transport data, thereby suggesting a close connection between the electronic and lattice degrees of freedom in this Sn-based quasi-skutterudite. The temperature dependence of $\rho(T)$ is analyzed in terms of a power-law for resistivity pertinent to Fermi liquid picture. Giant, positive magnetoresistance (MR) $\approx$ 80$\%$ is observed in Gd$_3$Ir$_4$Sn$_{13}$ at 2~K with the application of 9~T. The giant MR and the double magnetic transition can be attributed to the quasi-cages and layered antiferromagnetic structure of Gd$_3$Ir$_4$Sn$_{13}$ vulnerable to structural distortions and/or dipolar or spin-reorientation effects. The giant value of MR observed in this class of 3:4:13 type alloys, especially in a Gd-compound, is the highlight of this work.
\end{abstract}
\maketitle
\section{\label{INTRO} Introduction}
\begin{figure}[!b]
\centering
\includegraphics[scale=0.09]{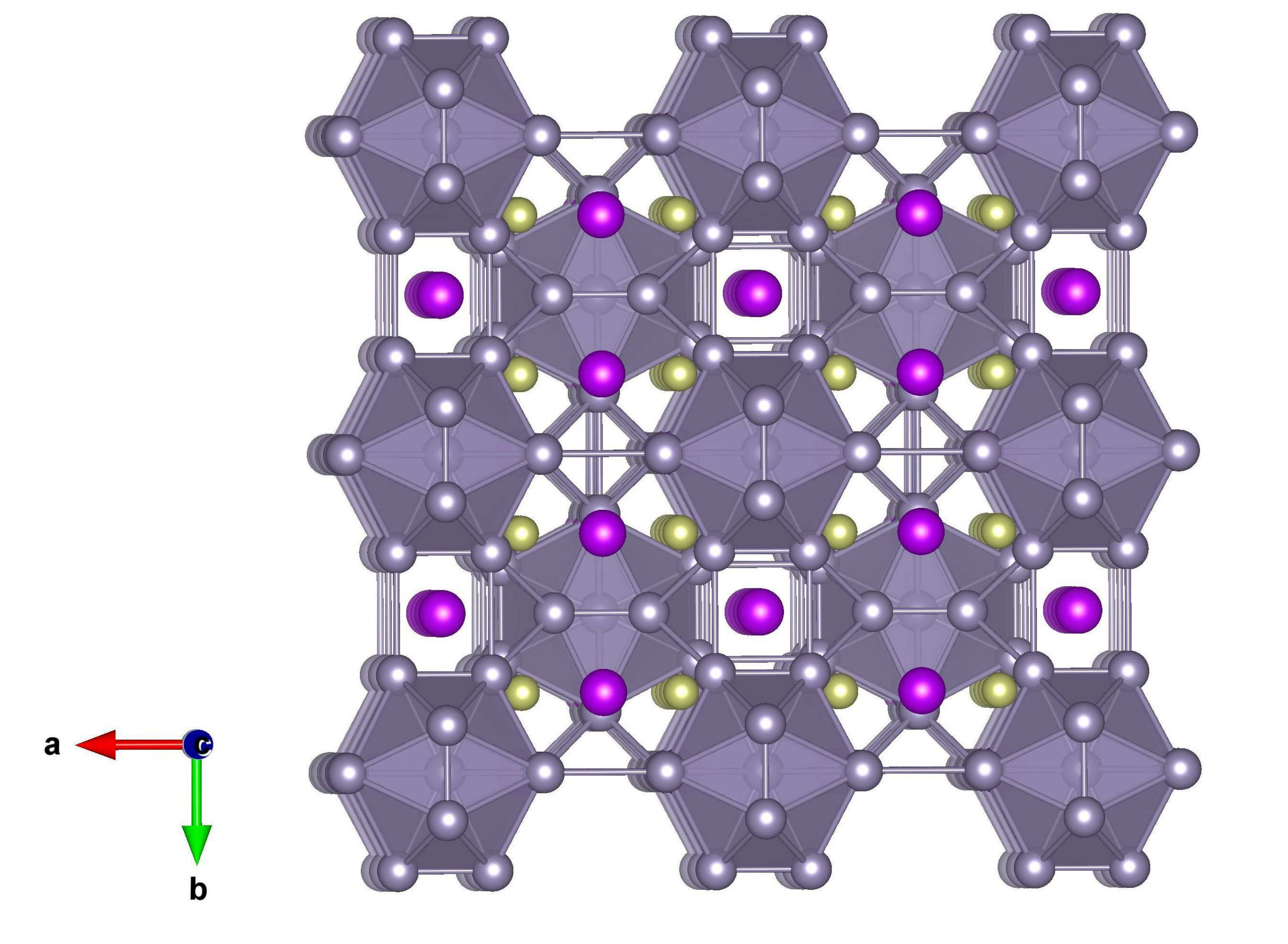}
\caption{\label{fig_str} (color online) The crystal structure of Gd$_3$Ir$_4$Sn$_{13}$ is depicted in the polyhedral coordination. The violet spheres are Gd, the yellow are Ir and the grey are Sn. }
\end{figure}

\indent
$R_3T_4X_{13}$ are caged structure compounds in the class of 
strongly correlated electron intermetallics where, 
$R$ is either a rare-earth element, an early $d$-block element 
such as Sc or Y, or the alkali-earth metals Ca or Sr.
$T$ stands for a Group VIII $d$-electron element, 
and $X$ is either In, Ge, or Sn. The $R_3T_4X_{13}$ 
compounds crystallize in
the cubic space group $Pm\overline{3}n$. Remeika {\em et al.}
\cite{remeika_ssc_34_1980new} 
were the first to report on $R_3T_4X_{13}$ phases but, the 
literature refers to the archetypal $R_3T_4X_{13}$ 
phase as either the Pr$_3$Rh$_4$Sn$_{13}$-structure type 
\cite{vandenberg_mrb_15_1980}, or the Yb$_3$Rh$_4$Sn$_{13}$-type 
\cite{hodeau_ssc_36_1980crystal}.  This structure type allows a 
single crystallographic site for occupation by each 
of the $R$ and $T$ atoms, and two distinct sites are available 
to the $X$ atom. A schematic of the crystal structure
is presented in Fig~\ref{fig_str} showing the cage-like
structural network. $R_3T_4X_{13}$ attracted attention 
mainly due to the discovery of superconductivity with superconducting 
temperatures as high as $T_{sc}$ = 8 K in 
Yb$_3$Rh$_4$Sn$_{13}$ and Ca$_3$Rh$_4$Sn$_{13}$
\cite{remeika_ssc_34_1980new} but this structure-type 
has been of enduring interest due to its amenability 
to many different elements and the wide 
variety of physical properties.
\cite{kase_physica_471_2011,kulkarni_prb_84_2011crossover} 
Remeika {\em et al.,} reported the crystal structure of $R_3T_4X_{13}$
in a cubic primitive unit cell,\cite{remeika_ssc_34_1980new} 
however, also suggested face-centered and tetragonal 
structures as possible unit cells. 
Hodeau {\it et al.,} reported a body-centered unit cell 
for the same class of compounds.\cite{hodeau_ssc_42_1982structural} 
Modifications to the cubic structure with changes in stoichiometry 
were suggested in Ref.[7].\cite{eisenmann_jlcm_1986cage} 
The first two reports on the crystal structure of 
the germanides are from Segre {\em et al.,}
\cite{segre_1981properties} and from Bruskov {\em et al.,}
\cite{bruskov_22_1986crystal} who proposed a disordered 
variant of the crystal structure where Ge atoms partially 
occupy two different 24$k$ sites. A positional disorder 
of the $X$ atom or the formation of random mixture of 
$R/T$ and $X$ atoms are known.
\cite{mudryk_jpcm_13_7391_2001physical,niepmann_znatur_2001structure} 
Doubling of the unit cell has been observed with a
non-centrosymmetric $I4_132$ space group in some 
of the stannides ($X$ = Sn) of $R_3T_4X_{13}$.
\cite{nagoshi_jpsj_75_2006magnetic,bordet_ssc_78_1991synchrotron}
\\
\indent
Stannides ($X$ = Sn) in this class have been investigated 
in some detail, for example, Ce$_3$Ir$_4$Sn$_{13}$
\cite{sato_physica_186_1993magnetic} and 
Ce$_3$Rh$_4$Sn$_{13}$\cite{kohler_jpcm_19_2007low} 
are classified as heavy-fermion systems where high 
effective mass or larger density of states at the Fermi
level have been observed. The stannide-$R_3T_4X_{13}$ 
display superconductivity in the case of $R_3$Rh$_4$Sn$_{13}$ 
(for $R$ = La, Yb, Ca, Sr, Th) while antiferromagnetism 
for Gd and Eu.\cite{hodeau_ssc_42_1982structural} 
A double-magnetic phase transition at 
0.6~K and 2.1~K is exhibited by 
the Ce-based stannide, Ce$_3$Ir$_4$Sn$_{13}$.
\cite{nagoshi2005anomalous} 
On the other hand, Eu$_3$Ir$_4$Sn$_{13}$ where
Eu shows valence fluctuation between non-magnetic 
Eu$^{3+}$ (4$f^6$) and magnetic Eu$^{2+}$ (4$f^7$),
shows an antiferromagnetic phase transition at 10~K.
\cite{mendoncca2006antiferromagnetic} 
Nagoshi {\em et al.,} studied single crystals of Gd$_3$Ir$_4$Sn$_{13}$
using magnetic susceptibility, resistivity, Hall effect and 
epithermal neutron scattering.\cite{nagoshi_jpsj_75_2006magnetic}
They identified a structural distortion characterized by the 
propagation vector (1/2, 1/2, 0). Only a single magnetic transition 
was identified at 10~K from macroscopic measurements. 
The magnetic moment direction of Gd in the ordered state 
is suggested to be perpendicular to the chain axis in each 
magnetic sublattice.\cite{nagoshi_jpsj_75_2006magnetic} 
Antiferromagnetic interactions in one-dimensional
chains together with ferromagnetic interaction between 
the nearest chains in the same sublattice was assumed to 
be the reason behind the phase transition at 10~K.
$R_3X_4$Sn$_{13}$ compounds have been, in fact, reported to display 
multiple phase transitions in close temperature intervals. 
For example, Eu$_3$Ir$_4$Sn$_{13}$ is reported to show two 
transitions in resistivity and specific heat occurring at 
$T^*\sim$ 57~K and at $T_N \approx$ 10~K.
\cite{mendoncca2015high,mendoncca2006antiferromagnetic}
The transition at $T^*$ was attributed to a structural 
distortion due to the displacement of Sn ions in the 
Sn(1)Sn(2)$_{12}$ cages while the transition at $T_N$ was 
attributed to antiferromagnetic ordering. 
Interestingly, the structural distortion and the 
antiferromagnetic ordering were characterized by the same 
propagation vector, $\bf q_c$ = (1/2, 1/2, 0).
\cite{mardegan2013structural}
Eventhough the structural transition at $T^*$ was observed 
to be suppressed by the application of external 
pressure up to 10.3~kbar, the magnetic transition 
was not affected.\cite{mendoncca2015high} 
With the application of magnetic field,
a new feature was observed to develop in the magnetically 
ordered state. Anomalous phase transitions were observed in 
Ce$_3$Ir$_4$Sn$_{13}$ as well, where the anomalies occurred 
at 0.6~K and at 2~K.\cite{nagoshi2005anomalous} 
The transition at 0.6~K was identified as antiferromagnetic 
whereas the one at 2~K was argued to be due to a change of 
band structure accompanied by a lattice expansion. 
Interestingly, magnetic susceptibility did not present
an anomaly at 2~K. It must be noted here that an earlier 
report on Ce$_3$Ir$_4$Sn$_{13}$ had shown that the 
2~K-transition split into a very sharp peak at 
2.10~K and a shoulder-like one at 2.18~K.
\cite{takayanagi1994two} A closely related compound, 
Eu$_3$Rh$_4$Sn$_{13}$, showed only one anomaly at 
11.2~K due to an antiferromagnetic transition
\cite{maurya2014detailed}
but, in the magnetization measured for $H \parallel$ [110] 
it showed two close-by transition when the applied field was 13~T. 
While the previous report\cite{mendoncca2006antiferromagnetic} 
speculated on the possibility of polaronic effects via electron-phonon 
coupling or Fermi surface induced changes in the conduction 
electrons scattering to be the origin of the transition at $T^*$,
present understanding\cite{mendoncca2015high} 
attributes the origin of the high 
temperature transition to structural disorder within the cages.
\\
\indent
In the present paper we endeavor to re-investigate 
Gd$_3$Ir$_4$Sn$_{13}$ using magnetization, specific heat 
and electrical resistivity
measurements on a polycrystalline sample. Even though a 
polycrystalline sample was used for the study, two very 
close-by magnetic transitions
in Gd$_3$Ir$_4$Sn$_{13}$ are identified through high resolution 
measurements. In addition, giant, positive MR  $\approx$ 80$\%$ is 
observed in this compound at 2~K which, to the best of our knowledge,
is the highest reported so far in this class of Gd-compounds.
\section{\label{EXP} Experimental details}
\indent
Polycrystalline samples of Gd$_3$Ir$_4$Sn$_{13}$ were prepared 
by arc melting the constituent elements Gd (99.99$\%$), 
Ir (-22 mesh, Premion 99.99$\%$ Alfa Aesar) and Sn 
(99.99$\%$ Alfa Aesar) together.  The elements weighed according to 
stoichiometric ratio were melted in the water-cooled 
Cu-hearth of Edmund Buehler arc melting furnace under 
static atmosphere of purified argon gas. A Zr-getter trap was 
used for purifying the Ar gas. The once-melted sample was 
flipped over and re-melted 5 times in order to ensure a single 
homogeneous phase. The melted samples were annealed for 
2 weeks at 680~$^\circ$C. 
Powder X ray diffractogram of pulverized sample of 
Gd$_3$Ir$_4$Sn$_{13}$ was obtained using a Philips X'pert diffractometer 
using Cu-K$\alpha$ radiation. Structural analysis of the x ray 
diffraction data was performed using Rietveld method\cite{rietveld} 
implemented in FullProf suite of programs.\cite{carvajal} 
Magnetic measurements were performed using a commercial Magnetic 
Property Measurement System (MPMS) and specific heat 
was measured using a commercial Physical Property 
Measurement System (PPMS) (both 
instruments from Quantum Design, San Diego). Electrical 
resistivity was measured on a bar-shaped 
sample of dimension 
$l \times b \times \phi$ $\approx$ 5~mm $\times$ 3~mm $\times$ 0.8~mm
using the ac transport option of the PPMS.
For magnetoresistance measurements, the magnetic field $B$, 
the current $I$, and sample length $l$ were arranged such that 
$B\perp I \parallel l$. Thermal conductivity and Seebeck coefficients were
measured on a bar shaped sample using the Thermal 
Transport Option (TTO) of the PPMS.
\section{\label{RES}Results}
\subsection{\label{str} Crystal structure $\&$ distortions}
\begin{figure}[!b]
\centering
\includegraphics[scale=0.55]{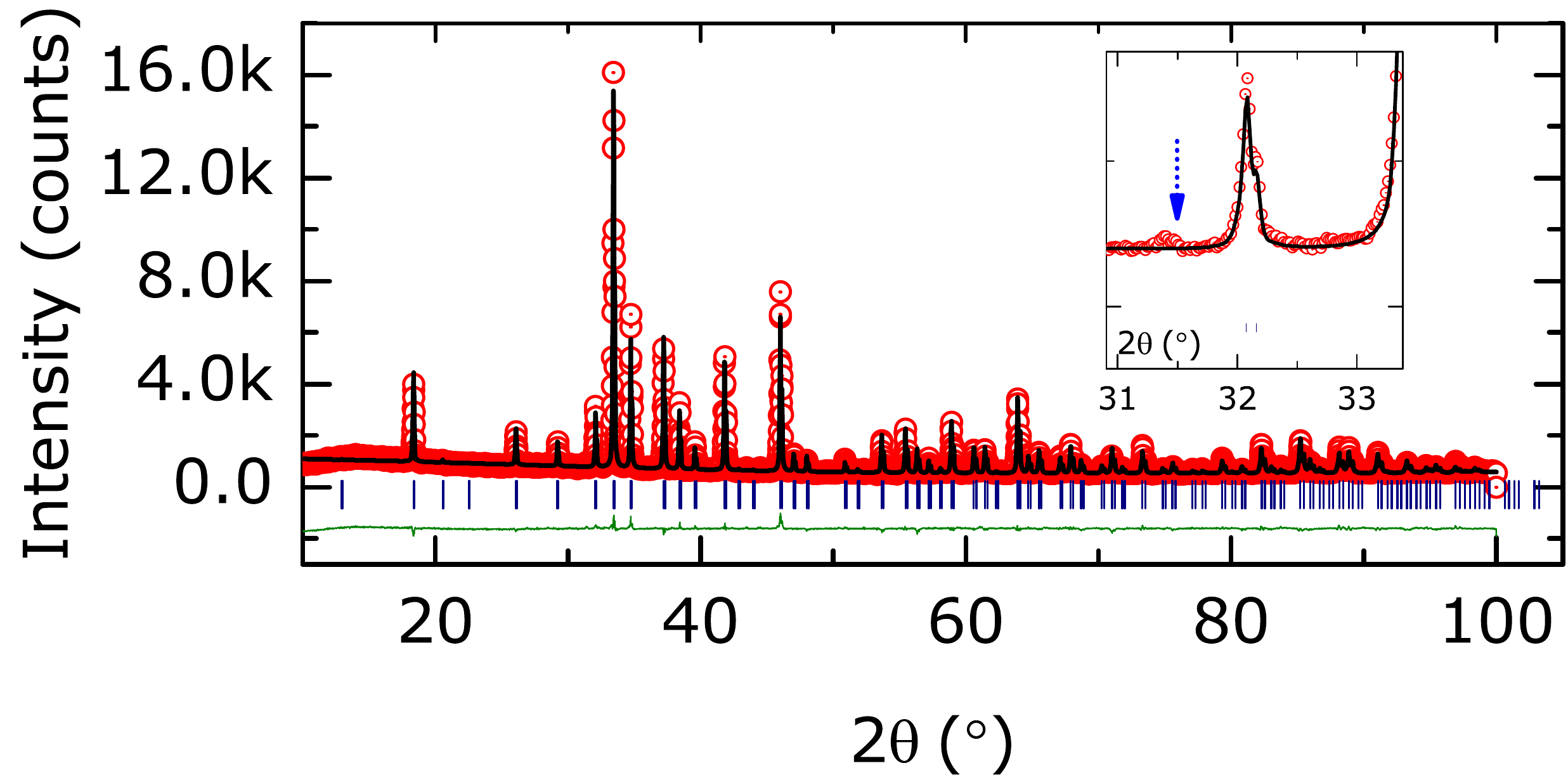}
\caption{\label{fig_xrd} (color online) The x ray diffraction pattern of Gd$_3$Ir$_4$Sn$_{13}$ is shown in red circles. The black solid line is the calculated pattern assuming $Pm\overline{3}n$ space group and the vertical bars are the allowed Bragg positions. The difference patterns is shown in green scatter. The inset of the graph highlights the region around 2$\theta$ = 31.5$^{\circ}$ where a superstructure peak is visible (the intensity axis is scaled down to show the feature clearly).}
\end{figure}

\indent
The crystal structure of $R_3T_4X_{13}$ compounds is generally 
described in cubic $Pm\overline{3}n$ space group 
($\#$223). Recent crystallography work on $R_3T_4X_{13}$
\cite{gumeniuk_dalton_41_20123}
compounds point towards the existence of subtle structural 
distortions in this class of compounds and confirms the 
findings of earlier reports.
\cite{remeika_ssc_34_1980new,vandenberg_mrb_15_1980,miraglia1986nature,hodeau_ssc_42_1982structural}
Rietveld refinement of the x ray data on Gd$_3$Ir$_4$Sn$_{13}$ 
was performed using the $Pm\overline{3}n$ space group. The results of the 
refinement are presented in Fig~\ref{fig_xrd} where the 
experimentally observed data are shown in circles and 
the calculated pattern 
as thick solid line. The refinement yielded a lattice 
parameter value of $a$ (\AA) = 9.6539(3). Nagoshi {\em et al.,}
\cite{nagoshi_jpsj_75_2006magnetic} pointed out the 
presence of several extra reflections mainly in the 
2$\theta$ range of 25 to 55$^{\circ}$ 
in the x ray diffraction pattern obtained on a single crystal of 
Gd$_3$Ir$_4$Sn$_{13}$. A close inspection of the x ray diffractogram
of our sample also revealed additional superstructure 
peaks signifying structural distortions in the present 
sample of Gd$_3$Ir$_4$Sn$_{13}$. 
The (2$\theta$, $(h, k, l)$) pairs of the peaks are as 
following: (31.5$^{\circ}$, $(1/2, 3/2, 3)$), (37.9$^{\circ}$, $(1/2, 7/2, 2)$),
(44.5$^{\circ}$, $(5/2, 7/2, 2)$) and (50.5$^{\circ}$, $(7/2, 7/2, 2)$). 
One such peak is shown enlarged in the inset of Fig~\ref{fig_xrd}
where the intensity is scaled down by a factor of 10$^{4}$. 
These peaks are indexed following the structural
distortion with a propagation vector $\bf q_c$ = (1/2, 1/2, 0) 
(a minor impurity belonging to $\beta$Sn
was observed at 43$^{\circ}$). The distorted structure of 
this type can be assigned to the space group $I4_132$,
similar to the superlattice structure for Gd$_3$Rh$_4$Sn$_{13}$ 
and La$_3$Rh$_4$Sn$_{13}$.\cite{miraglia1986nature} 
\subsection{\label{cp} Specific heat}
\begin{figure}[!b]
\centering
\includegraphics[scale=0.45]{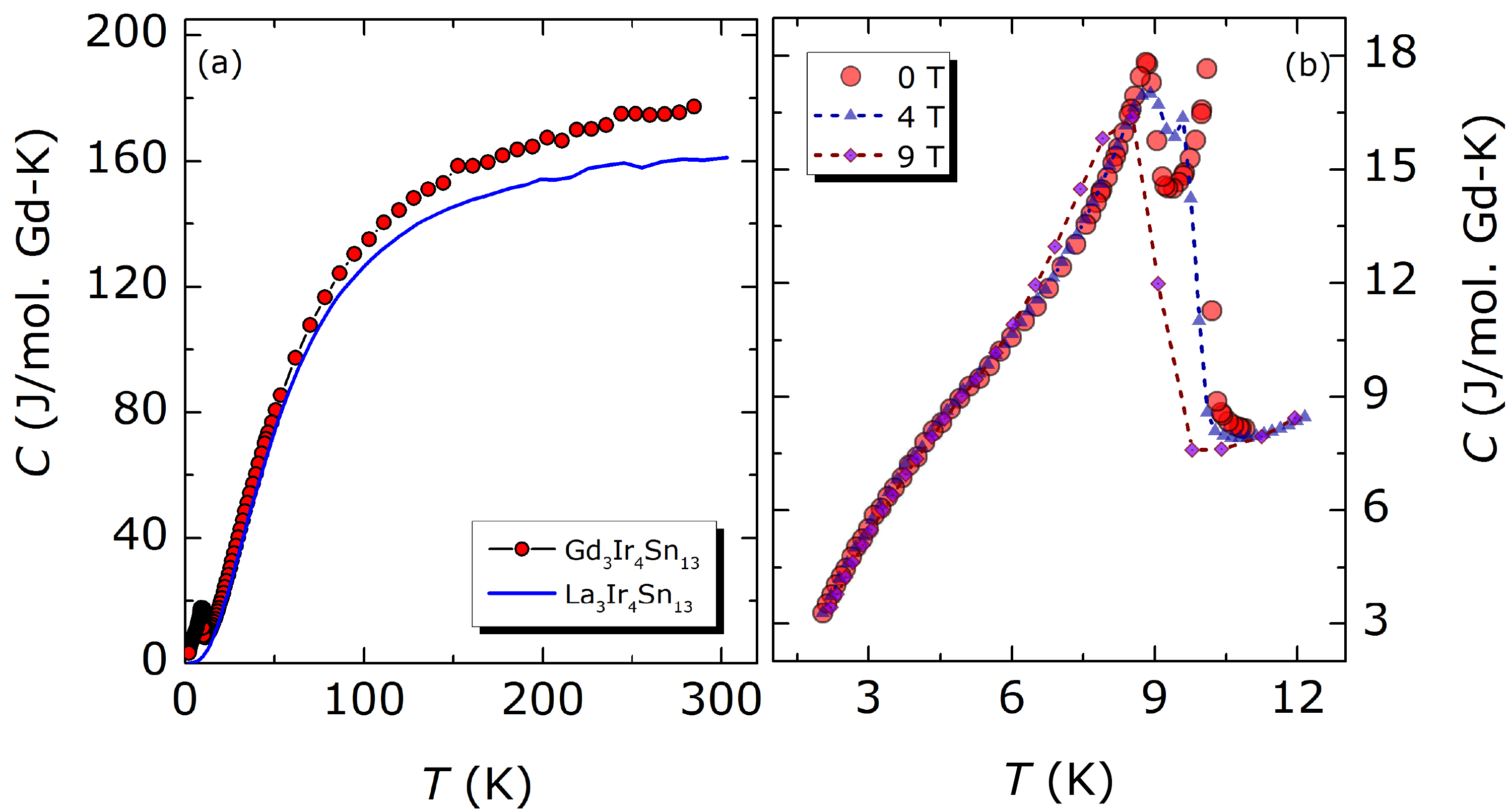}
\caption{\label{fig_cp} (color online) (a) The experimentally measured specific heat of Gd$_3$Ir$_4$Sn$_{13}$ plotted in circles. The data for non-magnetic analogue La$_3$Ir$_4$Sn$_{13}$ plotted as a line. The double-phase transition occurring at $T_{N1} \approx$ 10~K and $T_{N2} \approx$ 8.8~K are shown magnified in (b) where the low-temperature specific heat in 0, 4 and 9~T are plotted together.}
\end{figure}

\indent
The experimentally measured specific heat, $C_p(T)$, of 
Gd$_3$Ir$_4$Sn$_{13}$ is presented in the panel (a) of 
Fig~\ref{fig_cp} along with that of the
non-magnetic analogue La$_3$Ir$_4$Sn$_{13}$ which is 
shown in the figure using a solid line. The Dulong-Petit 
value of 3$nR \approx$ 
166~J/mol.Gd-K is recovered for Gd$_3$Ir$_4$Sn$_{13}$ at 
300~K ($n$ is 1/3 of the number of atoms
in the formula unit and $R$ is the universal gas constant). 
The $C_p(T)$ of Gd$_3$Ir$_4$Sn$_{13}$ resembles that
of other $R_3T_4X_{13}$ compounds reported in the literature.
\cite{strydom_jpcm_19_2007thermal,strydom_physica_403_2008r3ir4ge13}
However, the low-temperature region of specific heat 
gives evidence of a double magnetic phase transition with two 
nearby peaks at $T_{N1}\approx$ 10~K and $T_{N2}
\approx$ 8.8~K. The inset of Fig~\ref{fig_cp} magnifies the 
temperature region 
between 1~K and 14~K to highlight the double-peaks. 
Presented in the panel (b) are the specific heat data 
under applied magnetic fields
of 4~T and 9~T. It is observed that the application of 
magnetic field shifts the peak at $T_{N1} \approx$ 
10~K to low temperatures. 
Previous report on the magnetic and transport properties 
of Gd$_3$Ir$_4$Sn$_{13}$ did not identify the signature of 
double magnetic phase transition in their data.
\cite{nagoshi_jpsj_75_2006magnetic} The magnetic 
entropy, $S_m(T)$, was estimated 
by subtracting the specific heat of non-magnetic 
analogous compound La$_3$Ir$_4$Sn$_{13}$ from that of
Gd$_3$Ir$_4$Sn$_{13}$. In this way, $S_m \sim$17.2~J/mol-K 
was obtained at $T_{N1}$ which is equal to $R$ln(8) for the full multiplet of 
Gd$^{3+}$ ($S_m \approx$16~J/mol-K, at $T_{N2}$ ).
\\
\subsection{\label{mag} Magnetization}
\begin{figure}[!t]
\centering
\includegraphics[ scale=0.45]{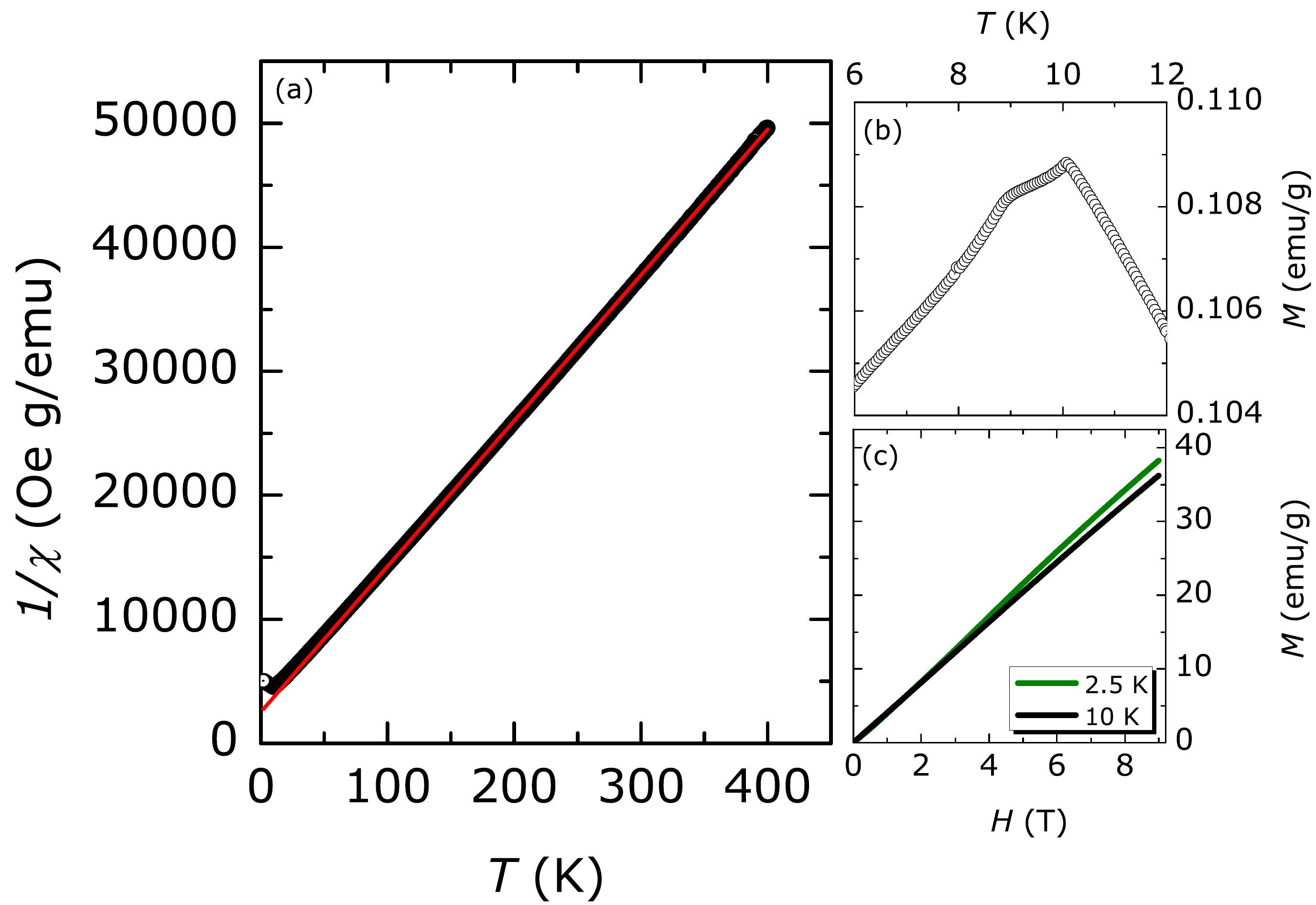}
\caption{\label{fig_mt} (color online) (a) The inverse magnetic susceptibility 1/$\chi(T)$ of Gd$_3$Ir$_4$Sn$_{13}$ plotted in circles along with the curve-fit using Curie-Weiss law shown as red solid line. (b) Shows the region of $M(T)$ displaying the double transitions at $T_{N1}\approx$10~K and $T_{N2}\approx$ 8.8~K. (c) Displays the magnetization isotherms at 2.5~K and 10~K which clearly shows the antiferromagnetic nature persisting up to 9~T.}
\end{figure}

\indent
The inverse magnetic susceptibility, 1/$\chi(T)$, of Gd$_3$Ir$_4$Sn$_{13}$ 
at 500~Oe is presented in the panel (a) of Fig~\ref{fig_mt}. 
A curve-fit assuming Curie-Weiss law performed in the range 80--300~K 
is shown as a straight line. The
Curie-Weiss fit leads to a value of effective paramagnetic 
moment $\mu_\mathrm{eff}$
= 7.2~$\mu_\mathrm{B}/$Gd-atom and Curie temperature 
$\theta_\mathrm{CW}$ = -21~K. The 
$\theta_\mathrm{CW}$ and $\mu_\mathrm{eff}$
values are comparable to the theoretical free-ion value for 
Gd$^{3+}$, $\approx$ 7.94~$\mu_\mathrm B/$ Gd-atom.
The observed deviation of the experimental value of magnetic moment
from that of the free ion Gd is interesting and might
hint at the contributions from $5d$ electrons.
The double-transition in Gd$_3$Ir$_4$Sn$_{13}$ is evident in 
magnetic response seen in the panel (b) of Fig~\ref{fig_mt}, 
where $M(T)$ in zero field cooled cycle measured at 500~Oe is 
presented in enlarged scales. The panel (c) of Fig~\ref{fig_mt}
shows the magnetization isotherms, $M(H)$ at 2.5~K and 
at 10~K plotted together. No signature of
ferromagnetism or metamagnetism is observed in the
magnetization isotherms while up to 9~T, the 
antiferromagnetic behaviour is retained.
\subsection{\label{resis} Resistivity and magnetoresistance}
\begin{figure}[!t]
\centering
\includegraphics[scale=0.35]{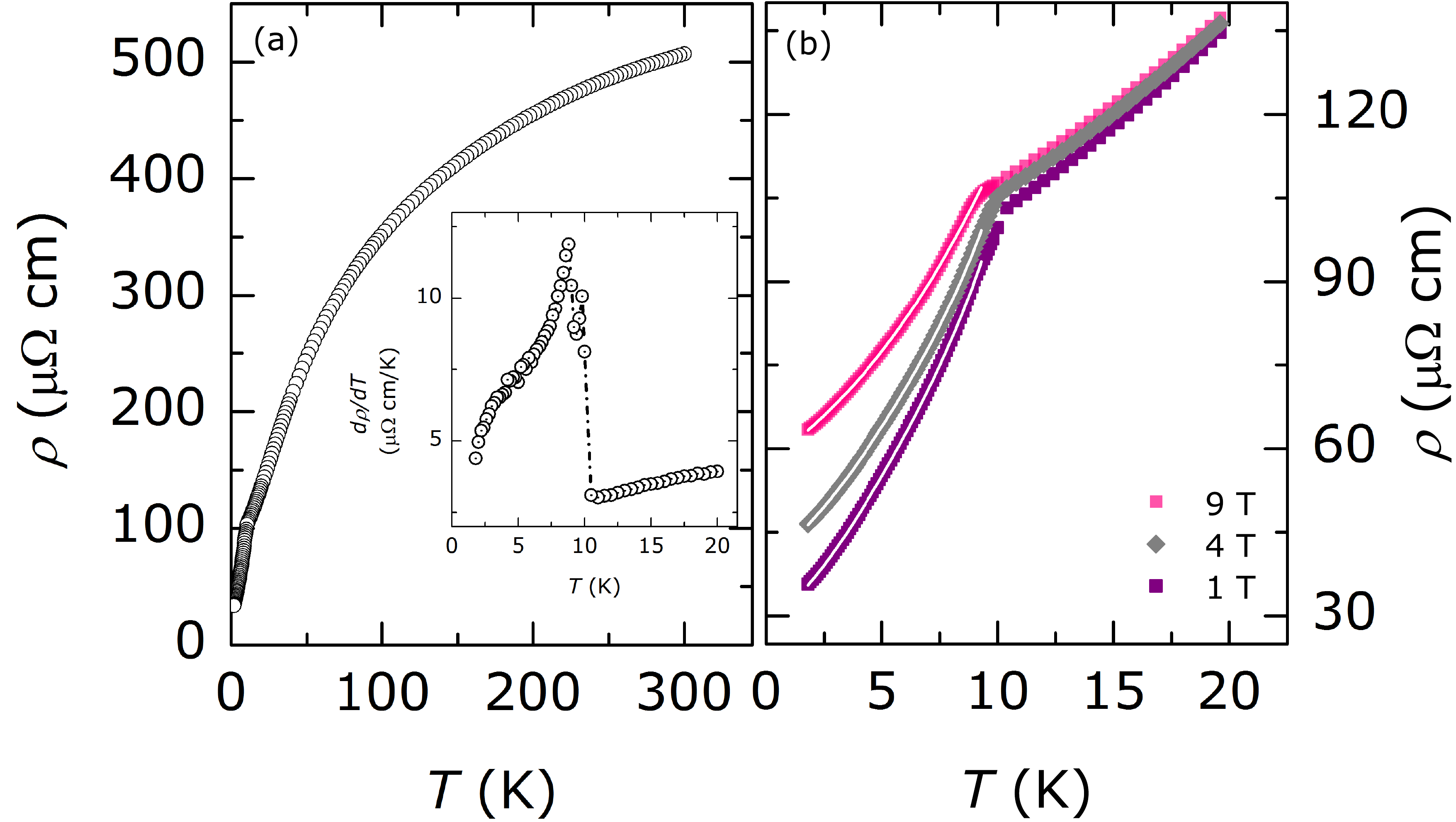}
\caption{\label{fig_rho} (color online) (a) The electrical resistivity, $\rho(T)$, of Gd$_3$Ir$_4$Sn$_{13}$ in zero applied field. The derivative plot d$\rho$/dT versus T is shown in the inset of (a). The double transitions are clearly observed in the derivative plot. Panel (b) shows the $\rho(T)$ obtained in applied fields 1, 4 and 9~T. The white solid lines are the curve fits according to Eqn~(\ref{eqn1}).}
\end{figure}

\begin{figure}[!t]
\centering
\includegraphics[scale=0.45]{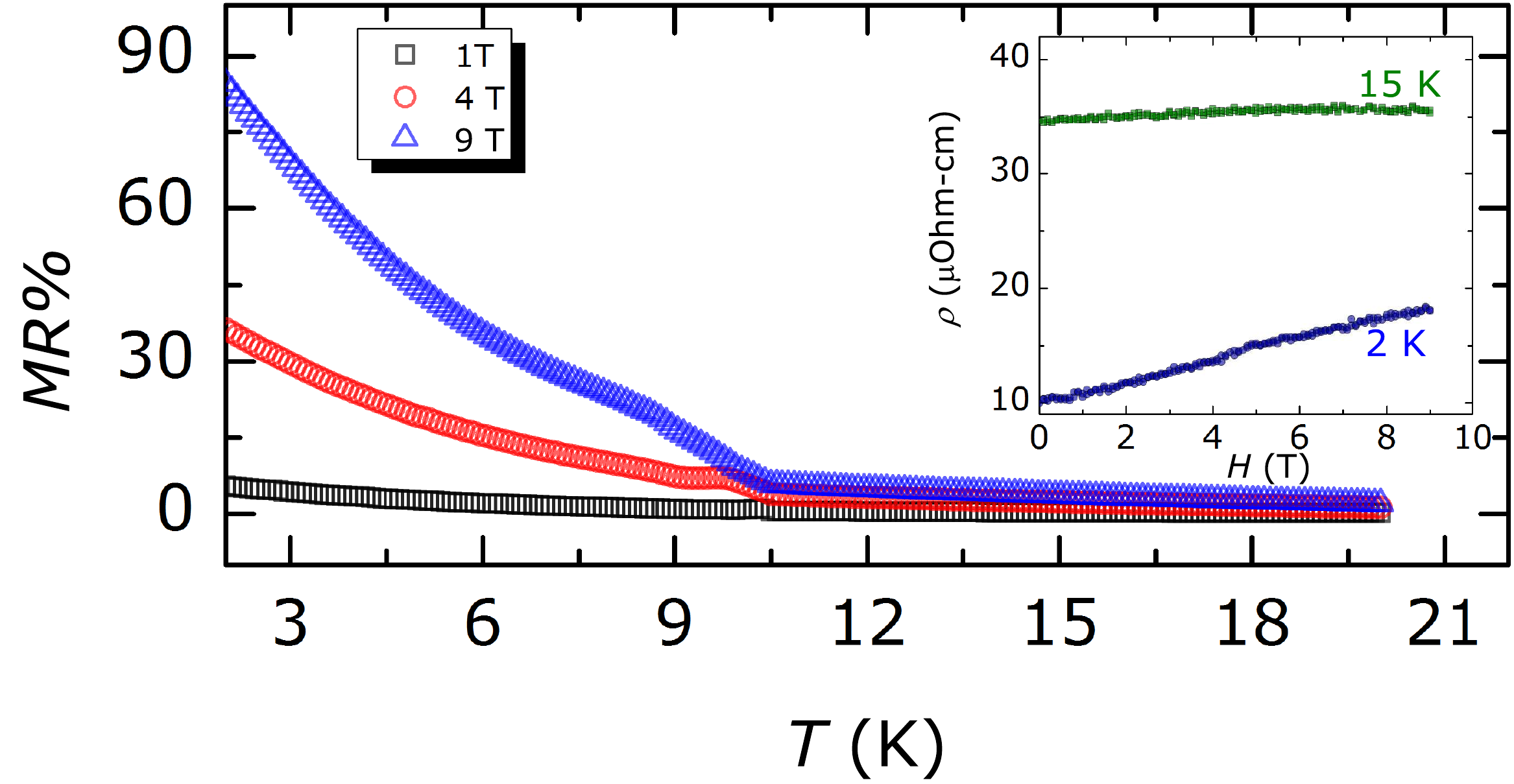}
\caption{\label{fig_mr} (color online) The magnetoresistance, MR $\%$, of Gd$_3$Ir$_4$Sn$_{13}$ for applied fields of 1, 4 and 9~T. The inset shows the isothermal magnetoresistance upto 9~T at 2~K as well as at 15~K. A giant, positive value of MR $\approx$ 80$\%$ is observed at 2~K upon the application of 9~T field.}
\end{figure}

\indent
The electrical resistivity, $\rho(T)$, of Gd$_3$Ir$_4$Sn$_{13}$ in 0~T 
is presented in the main panel of Fig~\ref{fig_rho}. In general, a 
metal-like behaviour is observed with a prominent anomaly at 
$\approx$ 10~K. This "kink" corresponds to $T_{N1}$. The residual
resistance ratio (RRR) defined as $\rho(300~K)/\rho(2~K)$ is 
approximately 50, which is higher than the value of RRR = 3 obtained 
on Gd$_3$Ir$_4$Sn$_{13}$ single crystals.
\cite{nagoshi_jpsj_75_2006magnetic} 
This difference between our data and that of 
Nagoshi {\em et al.,} is large only in the magnetic state.
Above the transition temperature the ratios are comparable.
The deviation from the RRR of a single crystal could 
possibly be due to a combination of grain boundary effects 
and electron-phonon scattering at high temperature. 
The Mott term also could be relevant because the variation 
becomes non-linear at relatively low temperature which is 
a signature of $s$-$d$ interband scattering. 
Enhanced quality of the polycrystalline samples  reflected 
in RRR and visibility of low-temperature
transition were observed in Ce$_3$Pd$_{20}$Si$_{6}$ 
for example.\cite{prokofiev2009crystal} A closer inspection 
of the temperature derivative d$\rho$/dT plotted in the 
inset (a) brings up both the anomalies $T_{N1}$ and $T_{N2}$ 
clearly. It is observed that the electrical resistivity of 
Gd$_3$Ir$_4$Sn$_{13}$ below the transition at $T_{N1}$ can 
be faithfully described by the expression;
\begin{equation}
\rho(T) = \rho_0 + AT^n
\label{eqn1}
\end{equation}
In this expression, the first term accounts for impurity 
scattering.
A power-law of the form $AT^n$ accounts for the normal Fermi-liquid
quasiparticle excitation where $1 < n < 2$. During the fit,
the exponent $n$ was left as a free parameter yielding a 
value of $n \approx$ 1.52(2) for
0~T while it gradually increased to 1.63(2) for 9~T. 
The parameters derived from the fit are collected in Table~\ref{tab1}.
\\
\indent
The magnetoresistance, defined as 
MR$\%$ = $\frac{\rho(H) - \rho(H=0)}{\rho(H=0)} \times 100$, 
of Gd$_3$Ir$_4$Sn$_{13}$ is plotted in the main 
panel of Fig~\ref{fig_mr}. The MR is calculated for $H$ = 1, 4 and 9~T. 
As can be seen from the figure, with the application of 9~T giant, 
positive MR $\approx$ 80$\%$ is obtained at 2~K. A progressive 
increase of MR with applied magnetic field can be observed. What is 
evident is the sudden increase of MR at $T_{N1}$ suggesting a 
close connection between the appearance of MR and the 
antiferromagnetic
transition. Above $T_{N1} \approx$ 10~K, the magnetoresistance 
is nearly zero for any value of applied field. In the inset of Fig~\ref{fig_mr}, 
the isothermal magnetoresistance is plotted for 2~K and for 15~K. 
The positive MR is clearly reproduced for the isotherm at 2~K whereas
at 15~K, a linear behaviour is recovered.
\subsection{\label{tto} Thermal transport}
\begin{figure}[!t]
\centering
\includegraphics[scale=0.53]{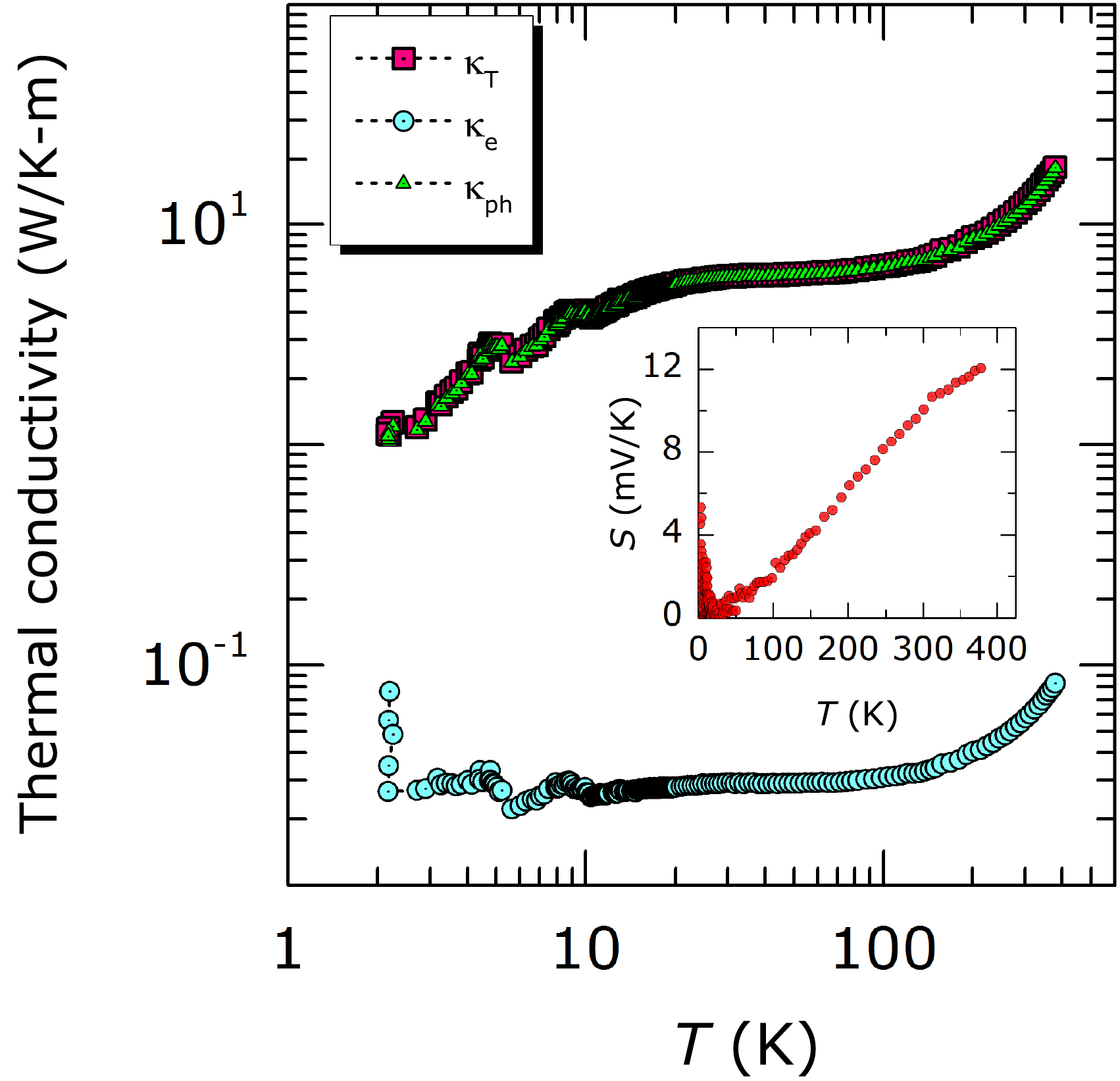}
\caption{\label{fig_tto} (color online) The total thermal conductivity $\kappa_\mathrm{T}$ of Gd$_3$Ir$_4$Sn$_{13}$ is plotted along with the electronic ($\kappa_\mathrm e$) and the phonon contributions ($\kappa_\mathrm{ph}$). Notably, the double-transition is evident in the thermal transport data also. The inset shows the Seebeck coefficient.}
\end{figure}

\indent
The experimentally measured thermal conductivity of 
Gd$_3$Ir$_4$Sn$_{13}$ is presented in the main panel of 
Fig~\ref{fig_tto} (a) denoted as $\kappa_\mathrm{T}$.
The electronic contribution to the total thermal conductivity 
is estimated using the Wiedemann -Franz law as 
$\kappa_\mathrm e = \frac{L_0T}{\rho(T)}$. 
The Lorenz number is 
$L_0$ = $\left(\frac{\pi k^2_B}{e \sqrt{3}}\right)^2$ = 2.54 
$\times$ 10$^{-8}$ W$\Omega K^{-2}$ and 
$\rho(T)$ is the electrical resistivity. The $\kappa_\mathrm e$ 
estimated this way is subtracted from the total thermal conductivity to obtain the
phonon contribution, $\kappa_\mathrm{ph}$ 
($\kappa_T$(T) =  $\kappa_\mathrm e(T)$ + $\kappa_\mathrm{ph}(T)$). 
Both $\kappa_\mathrm e$ and 
$\kappa_\mathrm{ph}$ are shown in Fig~\ref{fig_tto}. 
It is interesting to note that the magnetic transition
at $T_{N1}$ is clearly reflected in the thermal conductivity data. 
The thermal conductivity data bears a resemblance to the thermal transport 
data of other quasi-skutterudite compounds like Y$_3$Ir$_4$Ge$_{13}$.
\cite{strydom_jpcm_19_2007thermal} However, the thermopower of
Gd$_3$Ir$_4$Sn$_{13}$ exhibits a behaviour slightly 
different from that of the Ge-based
compounds of this class.
\cite{strydom_jpcm_19_2007thermal,strydom_physica_403_2008r3ir4ge13,kohler_jpcm_19_2007low}
A plateau-like region in the thermopower seen in the 
Ge-based compounds is not seen in the present case, instead, 
a linear increase of thermopower is observed. 
\section{\label{DISC} Discussion}
\indent
The phase transitions $T^*$ and $T_N$ reported to be observed in the stannides 
were not as close-by in temperature as has been observed 
as $T_{N1}$ and $T_{N2}$ in
Gd$_3$Ir$_4$Sn$_{13}$ ({\em i.e.,} at 8.8~K and 10~K). The 
double-transition in the present compound is observed in 
$M(T)$, $C_P(T)$, $\rho(T)$ and in thermal conductivity and 
hence is a intrinsic bulk effect. At lower applied fields like 500~Oe, 
both the transitions $T_{N1}$ and $T_{N2}$ are clearly observed in 
the $M(T)$ curve (see Fig~\ref{fig_mt}). When the applied 
field increases, the transition at $T_{N2}$ becomes less 
conspicuous in $M(T)$, but is still clearly identified 
in d$M(T)$/dT. It is noted that with 
applied field, the $T_{N1}$ shifts to lower temperature
while that at $T_{N2}$ remains unaffected. Such a feature is 
also reflected in the specific heat. The earlier report on single crystals of 
Gd$_3$Ir$_4$Sn$_{13}$ proposed a chain-like magnetic 
structure for this compound with interlayer antiferromagnetic 
coupling and weak ferromagnetism between the layers.
\cite{nagoshi_jpsj_75_2006magnetic} In order to test 
whether any presence of
ferromagnetism can be detected, we performed ac 
susceptibility measurements (data not shown here). 
Though signs of both $T_{N1}$ and $T_{N2}$ were clear 
in the data, there was no frequency dispersion present.
Also, the imaginary part of the ac susceptibility did not 
show any temperature dependence ruling out any 
dissipative or ferromagnetic 
terms. The type of antiferromagnetic order in Gd-alloys 
can be inferred from scrutinizing the specific heat.
\cite{bouvier1991specific,blanco1991specific}
The peak value of magnetic specific heat at the transition 
temperature is estimated to be 20.15~J/mol~K if the 
magnetic structure is 
any equal moment structure like simple antiferro-, ferro- or 
heli-magnetic. Any amplitude modulation is 
expected to reduce this value.
In the present case of Gd$_3$Ir$_4$Sn$_{13}$, a non-collinear 
magnetic structure may be expected following this test. 
Amplitude modulated magnetic structures are very common 
among highly anisotropic rare earth compounds with orbital magnetism.
However, Gd is a $S$-state ion with $L$ = 0 and hence one 
could assume that dipolar interactions can become relevant as has been 
reported in several Gd-compounds.\cite{rotter2003dipole}
It is interesting to note also that the $J$-$J$ coupling scheme 
becomes relevant in the case
where Hund's coupling is modest compared to the spin-orbit 
coupling and it was shown by Niikura {\em et al.,} that one of the 
seven $4f$ electrons of Gd can result in carrying a 
quadrupole moment.\cite{niikura2012quadrupole}
\\
\indent
The observation of positive giant magnetoresistance in 
Gd$_3$Ir$_4$Sn$_{13}$, is an important finding of the present work. 
The $R_3X_4$Sn$_{13}$ compounds are not reported to 
exhibit giant or positive magnetoresistance. 
However, magnetoresistance anomalies in Gd-alloys 
have been studied in detail and documented well.
\cite{sampathkumaran1995magnetoresistance,mallik1998magnetic,morellon1998giant,morellon2001giant,mazumdar1996positive,mazumdar1999low}
It is interesting to note that the temperature dependence of electrical resistivity
of Gd-alloys like GdCu$_2$Si$_2$  and GdNi$_2$Si$_2$
resemble that of Gd$_3$Ir$_4$Sn$_{13}$.
\cite{sampathkumaran1995magnetoresistance} 
The MR is observed to be positive 
for GdCu$_2$Si$_2$ above the N\'{e}el 
temperature.\cite{sampathkumaran1995magnetoresistance}
This observation suggested that the contribution
from the paramagnetic Gd fluctuation is small
compared to that from the influence of magnetic field
on the Fermi surface. On the other hand, for the Ni-based
alloy, the MR is negative in the paramagnetic state. Though
this feature is reminiscent of Kondo spin fluctuation phenomena,
since Gd $f$ electrons are well-localized, this feature is
attributed to the spin fluctuations from the Ni $d$-band.
In contrast to these two features, the MR  of 
Gd$_3$Ir$_4$Sn$_{13}$ is temperature 
independent at 15~K which is above $T_{N1}$. From the analysis
of effective paramagnetic moment, it was seen that possible
contributions from the 5$d$ electrons are present in
Gd$_3$Ir$_4$Sn$_{13}$. 
Gd$_2$Ni$_3$Si$_5$ is another compound that is reported to 
exhibit positive giant magnetoresistance.
\cite{mazumdar1996positive} However, the MR was not 
correlated to the rare earth moments but to the magnetic
ordering of the lattice as the magnitude of MR is
found to be significantly reduced in the paramagnetic state.
Notably, the MR was observed to be large in 
non-magnetic rare earths other than 
Gd in $R_2$Ni$_3$Si$_5$ (Gd$_2$Ni$_3$Si$_5$ showed $\approx$ 12$\%$ at 4.4~K 
for 45~kOe).\cite{mazumdar1996positive}
In Gd$_3$Ir$_4$Sn$_{13}$ also the enhancement of MR
is prominent below the magnetic ordering temperature
which hints at the connection between magnetic ordering
and giant MR. Similar to Gd$_3$Ir$_4$Sn$_{13}$, Gd$_2$Ni$_3$Si$_5$
also has a layered magnetic structure and this structural
feature also might play a role in the high value of
MR. 
On the other hand, giant magnetoresistance observed in Gd-based 
magnetocaloric alloys like Gd$_5$(Si$_{1.8}$Ge$_{2.2}$)
\cite{morellon1998giant} is attributed to the presence of
first-order magnetostructural transition. Note that in 
this compound a negative MR was observed. 
A similar feature was observed in Ge-rich compound 
Gd$_5$(Si$_{0.1}$Ge$_{0.9}$)$_4$.\cite{morellon2001giant} 
\\
\indent
Usually, positive MR can arise from Lorentz contribution 
to resistivity. In pure metals, high cyclotron frequency 
and relaxation times can lead to positive MR. 
However, such a scenario does not 
hold in the case of Gd$_3$Ir$_4$Sn$_{13}$ which shows a higher
low temperature resistivity than pure metals. Another 
possibility is for positive MR in antiferromagnets to arise 
due to enhancement of spin fluctuations.
\cite{yamada1973magnetoresistance} 
However, MR in such case is not as high as been observed 
in Gd$_3$Ir$_4$Sn$_{13}$. Gd-ion being an $S$-state ion, 
crystalline electric field effects can be ruled out 
as reasons for the observed giant MR. At this point 
it is interesting to mention that the discovery of 
giant magnetoresistance was made on layered magnetic 
structures with antiferromagnetic interlayer exchange.
\cite{binasch1989enhanced} A highly plausible reason for the 
double-magnetic transition and the positive 
giant magnetoresistance could be spin-reorientation 
effects taking place in the layered magnetic 
structure of Gd$_3$Ir$_4$Sn$_{13}$ with interlayer 
antiferromagnetic exchange.
\cite{nagoshi_jpsj_75_2006magnetic} However, it must 
be noted that spin-reorientation effects in 
Tb$_2$Ni$_3$Si$_5$ was evident in isothermal 
magnetization plots at 6~K and consequently was argued 
to have lead to positive giant MR.\cite{mazumdar1996positive} 
The magnetization
isotherms of Gd$_3$Ir$_4$Sn$_{13}$ at 2.5~K (Fig~\ref{fig_mt} (b)) 
do not present any deviation from linearity to support 
effects of spin-reorientation.
\section{\label{CONC} Conclusions}
The double-phase transition in the quasi-skutterudite Gd$_3$Ir$_4$Sn$_{13}$ is unambiguously identified through magnetization, specific heat, electrical resistivity, and thermal conductivity measurements. The two transitions occur very close to each other at $T_{N1}\approx$ 10~K and at $T_{N2}\approx$ 8.8~K. The transition at $T_{N1}$ is seen to shift to lower temperature with application of magnetic field hence, revealing the antiferromagnetic nature whereas the transition at 8.8~K is very robust and does not change upon applied field. The interesting result is the observation of positive giant magnetoresiatance of about 80$\%$ below $T_{N1}$. The layered quasi-1D magnetic structure of Gd$_3$Ir$_4$Sn$_{13}$ and/or dipolar interactions commonly found in Gd-based antiferromagnets could be the reason for the double-magnetic transition and the positive giant magnetoresisatnce.
\\ \\
$\ddagger$Present address: Department of Applied Physics, Birla Institute of Technology, Mesra, Ranchi, Jharkhand, India\\ \\
HSN and RKK acknowledge FRC/URC of UJ for postdoctoral fellowship. AMS thanks the SA NRF (93549) and UJ URC/FRC for financial assistance.
%
%
%
%
%
%
%

%
%
%
%
%
\clearpage\newpage
\begin{table}[!b]
\setlength{\tabcolsep}{0.4cm}
\caption{\label{tab1} The parameters extracted from the curve-fit to electrical resistivity $\rho(T)$ of Gd$_3$Ir$_4$Sn$_{13}$ under different values of applied magnetic fields, $H_\mathrm{app}$, assuming Eqn~(\ref{eqn1}) for resistivity behaviour under Fermi liquid picture.}
\begin{tabular}{llllll}\hline
$H_\mathrm{app}$ (Tesla)           &     0     & 1     & 4       & 9  \\ \hline
$\rho_0$ ($\mu\Omega$cm)           &     28.9    & 31.1    & 42.7      & 60.6  \\                                                             
$A$ ($\mu\Omega$cm/ K$^2$)         &     2.12   & 1.99   & 1.63       & 1.20  \\
$n$                                &     1.52  & 1.53  & 1.58    & 1.63  \\ \hline

\end{tabular}
\end{table}
\clearpage\newpage
\clearpage\newpage
\clearpage\newpage
\clearpage\newpage
\clearpage\newpage
\clearpage\newpage
\clearpage\newpage
\end{document}